\documentstyle[12pt,aasms4]{article}

\lefthead{Cunha et al.}
\righthead{Fluorine Abundances versus Metallicity}

\begin{document}

\title{Fluorine Abundances in the Large Magellanic Cloud and $\omega$ Centauri:
Evidence for Neutrino Nucleosynthesis?}  

\author{Katia Cunha}
\affil{Observat\'orio Nacional, Rua General Jos\'e Cristino 77, 20921-400,
S\~ao Cristov\~ao, Rio de Janeiro, Brazil; katia@on.br}

\author{Verne V. Smith}
\affil{Department of Physics, University of Texas El Paso, El Paso, TX 79968 
USA; verne@barium.physics.utep.edu}

\author{David L. Lambert}
\affil{Department of Astronomy, University of Texas at Austin, Austin, 
TX 78712, USA; dll@astro.as.utexas.edu}

\author{Kenneth H. Hinkle}
\affil{National Optical Astronomy Observatory, P.O. Box 26732, Tucson,
AZ  85726, USA; khinkle@noao.edu}

\begin{abstract}

The behavior of fluorine with metallicity has not yet been probed
in any stellar population. 
In this work, we present the first  fluorine abundances
measured outside of the Milky Way from a sample of
red giants in the Large 
Magellanic Cloud (LMC), as well the Galactic globular
cluster $\omega$ Centauri.  The fluorine abundances are derived from
vibration-rotation transitions of HF using infrared spectra obtained 
with the Phoenix spectrograph on the Gemini South 8.1m telescope.
It is found that the abundance ratio of F/O declines as the oxygen
abundance decreases.  The values of F/O are especially low in the two
$\omega$ Cen giants; this very low value of F/O probably indicates
that $^{19}$F synthesis in asymptotic giant branch (AGB) stars is not
the dominant source of fluorine in stellar populations.  The observed
decline in F/O with lower O abundances is in qualitative agreement
with what is expected if $^{19}$F is produced via H- and He-burning
sequences in very massive stars, with this fluorine then ejected in
high mass-loss rate Wolf-Rayet winds.
A quantitative comparison of
observations with this process awaits results from more detailed
chemical evolution models incorporating the yields from Wolf-Rayet
winds.  Perhaps of more significance is the quantitative agreement 
between the Galactic and LMC results with predictions from models
in which $^{19}$F is produced from
neutrino nucleosynthesis
during core collpase in supernovae of Type II.
The very low values of F/O in $\omega$ Cen are also in agreement with
neutrino nucleosynthesis models if the ``peculiar'' star formation
history of $\omega$ Cen, with 2-4 distinct episodes of star formation,
is considered.   

\end{abstract}


\clearpage

\section{Introduction}

Unlike its neighbors in the periodic table, the nucleosynthetic
origins of fluorine have remained somewhat obscure.  This
obscurity reflects a combination of theoretical and observational 
unknowns.
The one stable isotope of fluorine,
$^{19}$F, is not easy to produce in stars
 as it is readily destroyed by either 
proton or $\alpha$ captures during many phases of stellar evolution.
From an observational point-of-view, fluorine is difficult to detect
spectroscopically.  The only atomic lines that might be studied are the
ground-state lines of F I, falling in the far ultraviolet (UV) near
950\AA.  The molecule HF is readily detectable in  cool stars through
vibration-rotation transitions that fall in the infrared near $\sim$2.3 $\mu$m. 
The HF pure rotational
transitions are in the observationally challenging sub-millimeter
region and are not easily exploited with current technology (Neufeld
et al. 1997).

Attempts to account for fragmentary data on 
the abundance of fluorine in the Galaxy have
focussed mostly on three proposed sources: 1) neutrino-induced
spallation of a proton from $^{20}$Ne following the core-collapse phase
of a massive-star supernova (Woosley \& Haxton 1988) -- this is referred 
to as the $\nu$-process
(Woosley et al. 1990), 2) synthesis during He-burning thermal pulses
on the asymptotic giant branch (AGB), as suggested by Jorissen, Smith,
\& Lambert (1992), or 3) production of $^{19}$F in the cores
of stars massive enough to be Wolf-Rayet stars at the beginning of their
He-burning phase (Meynet \& Arnould 2000). 
The relative importance of the three sources is undetermined
theoretically and observationally. Indeed, it is not known if the
leading contributor of fluorine is among the listed trio.
Clues to the site of fluorine synthesis lie in the run of the fluorine
abundance with metallicity in the Galaxy, and the fluorine
abundances in different stellar populations.

To date, the only fluorine abundances measured in stars other
than the Sun were provided by
Jorissen et al. (1992), who 
 analyzed a few Galactic K and M giants, and a larger
number of $s$-process enriched Galactic giants (spectral types MS, S, and N),
but all were stars of near-solar metallicity.
It may be assumed that the fluorine abundance of the K and M
giants is unaffected by stellar evolution.
To expand knowledge of fluorine  abundances,
we present fluorine abundance measurements in two stellar
systems other than the Galaxy. We discuss red giants
of the Large Magellanic Cloud (LMC),  and
two red giants in the Galactic globular cluster $\omega$ Centauri.
These red giants  are significantly more metal poor than
the K and M giants analyzed by Jorissen et al. (1992)
and provide 
information for the first time on how the fluorine abundance varies
as a function of metallicity.

\section{Observations}

Fluorine abundances are derived in this
study  from high-resolution IR spectra obtained on the 8.1m
Gemini South telescope with the NOAO Phoenix spectrometer (Hinkle et al.
1998).  
A discussion of the observations and data reduction has appeared 
already in Smith et al. (2002) and details can be found there.
Spectra pertinent to the present analysis are single-order echelle spectra 
imaged on a 1024 x 1024 InSb Aladdin II array and 
were obtained with an entrance slit
defining the resolution, where R=$\lambda$/$\Delta$$\lambda$= 50,000 
($\sim$ 4-pixels for Phoenix on Gemini South).  The data were reduced to
one-dimensional spectra using IRAF.  

Spectra
analyzed here were centered near a wavelength of 23400\AA\ and
cover a 120\AA\ window. The HF 1--0 R9 line is included
along with CO first-overtone vibration-rotation lines (Smith et al.
2002).

\section{Analysis}

\subsection{Determining the Fluorine Abundances}

The stars observed for the HF line
include 9 red giant members of the LMC and 2 red giant members
of the Galactic globular cluster $\omega$ Cen.  All  of the stars
were analyzed recently -- see  Smith et al. (2002) for the
LMC giants, and Smith et al. (2000) for the $\omega$ Cen stars.
In both studies, a combination of photometry (mostly IR) and 
high-resolution spectroscopy is used to derive the stellar parameters
necessary for an abundance analysis: effective temperature (T$_{\rm eff}$),
surface gravity (parameterized as log g), microturbulent velocity
($\xi$), and stellar metallicity.  The same analysis techniques are
used in both the LMC and $\omega$ Cen studies.
We adopt the  stellar parameters  derived
in the Smith et al. (2000, 2002) papers.  In Table 1, we list the
program stars, along with published values of T$_{\rm eff}$, log g,
microturbulence, as well as Fe and O abundances.  The Fe and O
abundances in the LMC giants are taken from Smith et al. (2002),
with the uncertainties listed being the standard deviations from the
average of the 3 Fe I and 4 OH lines used in the abundance determinations.
The abundances of Fe and O for the $\omega$ Cen members and 
$\alpha$ Boo were taken from Smith et al. (2000).  In this paper
we adopt the spectroscopic notations for presenting abundances, with
A(x)= log[N(x)/N(H)] + 12.0, and
[x/y]= log[N(x)/N(y)]$_{\rm Program Star}$ -
log[N(x)/N(y)]$_{\rm Standard}$.

Fluorine abundances  listed in the last column of Table 1
are derived from spectrum synthesis of the HF 1--0 
R9 line. We  used a recent version of the LTE synthesis code 
MOOG (Sneden 1973). 
The model atmospheres
employed in the abundance analyses are based on two different versions
of the atmosphere
code MARCS (Gustafsson et al. 1975).
The LMC models were constructed with the SOSMARCS version (Plez, Brett,
\& Nordlund 1992; Edvardsson et al. 1993; Asplund et al. 1997),
while the $\omega$ Cen giants use an older version of the original
MARCS code.  Cunha et al. (2002)
compared synthetic spectra
generated from both SOSMARCS and MARCS models and found no significant
differences (in fact, almost unmeasurable differences) between the 
codes for temperatures in the range of the program stars studied here. 

In addition to the metal-poor giants from the LMC and $\omega$ Cen,
the nearby red giant $\alpha$ Boo was analyzed using the new IR spectral
atlas from Hinkle, Wallace, \& Livingston (1995) .  Although 
$\alpha$ Boo was included 
in the fluorine work by Jorissen et al. (1992), their analysis was
based on an earlier IR spectrum that did not have as high a S/N ratio
as the Hinkle et al. (1995) atlas.  Stellar parameters for $\alpha$ Boo
(Table 1) are from the recent analyses of
Smith et al. (2000, 2002).

Sample comparisons of real  with synthetic spectra are shown
in Figure 1 for star LMC 1.27  and in Figure 2 for ROA 324
 from $\omega$ Cen. 
These figures illustrate the
quality of the observed spectra and obtained fits.  Note in the case 
of ROA 324 that HF is not detected. The lowest abundance shown
in the figure is A(F)=3.10 and we adopt this value as a conservative
upper limit to the fluorine abundance (the line-profile depths at
this abundance are just slightly greater than the noise level in the
observed spectrum). 

The sensitivity of the fluorine abundance to input stellar parameters
was quantified by individually varying T$_{\rm eff}$, surface gravity, and
microturbulence to measure their separate effects on the derived
abundances.  A baseline model with T$_{\rm eff}$= 3600K, log g= 0.50,
and $\xi$= 2.5 km s$^{-1}$ was used.  The following abundance differences
were found for a R9 line having the same strength as in LMC 2.1158:
$\Delta$T$_{\rm eff}$= +70K gives $\Delta$A(F)= +0.14 dex,
$\Delta$(log g)= +0.3 dex gives $\Delta$A(F)= +0.02 dex, and
$\Delta$$\xi$= +0.5 km s$^{-1}$ gives $\Delta$A(F)= -0.03 dex.
These are typical stellar parameter uncertainties, and a quadrature
sum of the abundance changes yields a combined uncertainty of 0.15 dex
in the derived $^{19}$F abundance due to the estimated errors.

\subsection{Reanalysis of the Older Fluorine Data} 

In order to establish the behavior of fluorine with metallicity and across
stellar populations, 
 we add 
the only other available fluorine abundance measurements in the
Galactic disk: these are the non-$s$-process enriched K and M giants
studied in Jorissen et al. (1992).
We do not include the s-process enriched AGB
stars of spectral types MS, S, or C because Jorissen et al. (1992)
found that $^{19}$F is synthesized during He-burning thermal pulses,
and this freshly produced fluorine is mixed to the surfaces of these
AGB stars.  As our primary motivation here is to follow the general
chemical evolution of fluorine with metallicity 
(using both Fe and O as metallicity indicators),
inclusion of the self-polluted $^{19}$F-rich AGB stars will obscure
any metallicity trend.

A re-analysis of the Jorissen et al. (1992) abundances for the M-giants
was conducted, as the original results were derived
from  model atmospheres
computed by Johnson, Bernat, \& Krupp (1980). 
A comparison of models computed from the MARCS code and
those of Johnson et al. (1980) reveals that, in the outer line-forming
layers, the latter are slightly hotter ($\sim$50K): 
this 
small difference has a small effect $(\sim$ 0.1 dex) on the derived
 fluorine abundances.
 The Jorissen
et al. (1992) stellar parameters for these stars were  taken from
the papers by Smith \& Lambert (1985, 1986, 1990) and, as these were
also based mostly on the IR photometric-T$_{\rm eff}$ scale, we adopt
these same parameters, but use MARCS model atmospheres to re-analyze
the HF equivalent widths published by Jorissen et al. (1992).  

Table 2 lists the K and M giants from Jorissen et al. (1992), along
with their parameters and our derived Fe, O, and F abundances.  Not only were
the fluorine abundances recomputed with MARCS models, but also the
iron and oxygen abundances using the published equivalent widths
of Fe I and OH lines from Smith \& Lambert (1985; 1986; 1990).
In these earlier studies, the Fe I gf-values were ``astrophysical'',
based upon an analysis of the K giant $\alpha$ Tau.
The new abundances are based upon   gf-values 
from the Kurucz \& Bell (1995) linelist.  The effect of different
model atmospheres and gf-values is small: the average differences,
in the sense of (new-old) is A(Fe)=+0.06 and A(O)=-0.10.  The field K giants 
HR 5563 and HR6705 have fluorine abundances derived from the HF 1--0 
R9 line, as in the LMC and $\omega$ Cen giants, while the cooler
field M giants had the R13, R14, R15, and R16 lines measured.  The
comparison between the K and M giant $^{19}$F abundances shows complete
agreement, indicating that we have a homogeneous set of fluorine
abundances for some 23 red giants across three
stellar populations - the solar neighborhood, $\omega$ Cen, and the
LMC.

\subsection{The Solar Fluorine Abundance}

Jorissen et al. (1992) pointed out that their
sample of normal K- and M-giants with near-solar
metallicities gave an average $^{19}$F abundance somewhat larger than
the solar abundance. The average abundance as discussed in Jorissen et al.
(1992) sample of giants was A(F)= 4.69. The solar abundance is  somewhat
uncertain:  Anders \& Grevesse (1989) quote a meteoritic abundance
of A(F)= 4.48$\pm$0.06, and a solar photospheric abundances (from HF
lines of sunspots) of 4.56$\pm$0.30.

Our reanalysis of the Jorissen et al. data presented in Table 2 yields
slightly different abundances.  Taking only those K and M giants with
oxygen abundances in the range of A(O)= 8.7 - 8.9 (with solar being
8.77, which is the value found using 1-d MARCS models as discussed
by Allende Prieto, Lambert \& Asplund 2000),
the average $^{19}$F abundance is found to be 4.65$\pm$0.07:
marginally closer to the solar values, but still larger.
The 8 disk giants
have an average oxygen abundance of 8.78--within 0.01 dex of solar.
In discussing stellar fluorine abundances, 
 we adopt A(F)= 4.55 as the value of $^{19}$F
for the Sun, thus there is still a small 0.10 dex offset between
the solar value and the average of the near-solar metallicity disk
giants.

\section{Discussion}

The measurements of fluorine abundances in three
different stellar populations with differing metallicities hold clues to the
origins of fluorine. It is recognized, of course, that
evolution of an elemental abundance in a stellar system
depends on factors other than simply the yields from the principal site of
nucleosynthesis for the element. Such factors include the star formation
history and the dispersal of stellar ejecta throughout a stellar population;
consideration of these additional processes are argued to be important
here when comparing fluorine abundances in the globular cluster 
$\omega$ Cen to the Milky Way and the LMC.

\subsection{The Behavior of Fluorine with Oxygen and Iron}

The first comparison of fluorine with oxygen and iron is shown in
Figure 3: the top panel is A(F) versus A(O) and the bottom panel is
A(F) versus A(Fe).  
Looking first at oxygen, it is clear from Figure 3 that the field
K- and M-giant abundances are very close to solar.  In addition,
the metal-poor Galactic giant $\alpha$ Boo, with A(O)= 8.38 (or
[O/H]= -0.39) has a near-solar F/O abundance ratio (represented by
the solid line).  Including the
LMC red giants in the picture, one finds that, in general, they
also scatter about the solar F/O line:  the mean difference and
standard deviation of the Galactic and LMC points about the solar
F/O line is +0.03 $\pm$ 0.17 dex. This scatter is very similar to
that expected from the errors in the analysis. The striking deviations
from the solar F/O line are the two $\omega$ Cen giants, with both
falling well below a solar F/O abundance ratio: the upper limit in
ROA 324 is a firm limit, as shown in Figure 2. As will be discussed
in Section 4.2.3, the star fomation history in $\omega$ Cen is very
different from that of the Milky Way and the LMC and will lead
to a simple explanation for the low fluorine abundances in this
system. 

The comparison between F and Fe in the bottom panel of Figure 3 shows
more scatter about a scaled solar line for the disk and LMC giants.
In this case, the mean difference and standard deviation is 
+0.02 $\pm$ 0.30 dex,
almost twice as large as the scatter found about the F/O line. 
Although one of the $\omega$ Cen giants has an F/Fe ratio close to solar,
the conservative upper limit for ROA 324 is still significantly subsolar 
in F/Fe.

Figure 4 shows the relation of [F/O] using oxygen as the metallicity
indicator.  The use of oxygen removes supernovae of Type Ia as a
contributing source, as iron can have SN Ia's as a dominant parent
in some stellar populations; none of the possible sites put forth
as $^{19}$F sources involve these types of supernovae. Inspection
of this figure reveals a gradual decrease in the average [F/O]
value when going from the near-solar metallicity Galactic stars
to the lower metallicity LMC giants and $\alpha$ Boo. 

The continous lines plotted in Figure 4 are predictions from chemical
evolution models by Timmes et al. (1995) and Alib\'es, Labay \& Canal
(2001) for the behavior of [F/O] with oxygen for the $\nu$-process
yields as given by Woosley \& Weaver (1995). In both models, the [F/O]
values are slightly subsolar at solar metallicity by approximately
-0.1 dex. As pointed out by Alib\'es et al. (2001), $^{19}$F production
is very sensitive to the neutrino fluxes and spectra, so the yield is
uncertain; a 0.1 dex offset from solar is probably not significant. 
Timmes et al.
(1995) show (in their Figure 10) the effects of factor 2 variations
in the mass of ejected fluorine, and we have used their mean curve.
We note that quantitative chemical evolution model results for F 
production during
the Wolf-Rayet phase are not available in the literature and thus are
not discussed here.

\subsection{The Pros and Cons of the Sites for the Chemical Evolution of
Fluorine}

As discussed in the Introduction, three main sites have been put forth
as possible net $^{19}$F producers: 1) operation of a neutron source
during He-burning thermal-pulses in AGB stars, 2) mass loss from
WR stars during their phase of He-burning, and, 3) the $\nu$-process 
in supernovae of Type II. In our discussion of the
behavior of fluorine as a function of metallicity, we use oxygen as
the 'appropriate' metallicity indicator.

\subsubsection {AGB Stars}

Jorissen et al. (1992) probed the conditions under which $^{19}$F
could be produced as a result of a combination of $\alpha$-captures
originating on $^{14}$N, a neutron source, and proton captures,
all occurring during thermal pulses in AGB stars.  Although they
could explain the enhanced fluorine abundances observed in the s-process
enriched and $^{12}$C-rich stars, they did not make detailed predictions
about the chemical evolution of $^{19}$F.  
Due to the longer timescales of AGB
evolution, Jorissen et al. (1992) did argue that fluorine would likely
follow the behavior of iron, which also has a long timescale
component arising from SN Ia's, thus [F/Fe] would be $\sim$0.  Examination
of Figures 3 shows that the scatter about the F/Fe line is almost
twice as large as the scatter about the F/O line, as discussed
previously. This suggests that perhaps AGB stars are not the
place to look for major fluorine production.  In this instance the
$\omega$ Cen giants enter the picture as potentially crucial
points, due to the large s-process abundances found in many of the
$\omega$ Cen members: a result usually interpreted as showing an
especially large degree of AGB enrichment contributing to the chemical
evolution within $\omega$ Cen.  In such a picture, the $\omega$ Cen
giants should exhibit enhanced fluorine abundances, but show completely
opposite behavior.

\subsubsection {Wolf-Rayet Stars}

In massive stars,
$^{19}$F can be produced during the pre-explosive nucleosynthesis
in the helium shell from reaction sequences not involving neutrinos, but
by a series of thermonuclear reactions.  The synthesis of $^{19}$F
by these reactions consists of two modes, one consisting of a CNO
H-burning sequence and the other consisting of a set of He-burning 
sequences.  In the CNO H-burning sequence, a series of proton
captures and $\beta$-decays, initiated on $^{14}$N, leads to a
finite abundance of $^{19}$F, with $^{19}$F(p,$\alpha$)$^{16}$O
being the destructive reaction.  During He-burning, the sequences to
produce $^{19}$F begin with
an $\alpha$-capture on $^{14}$N and conclude with an $\alpha$-capture
on $^{15}$N ($^{15}$N($\alpha$,$\gamma$)$^{19}$F); this series of reactions
also involves neutron and proton captures, with the neutrons and
protons supplied by $^{13}$C($\alpha$,n)$^{16}$O and
$^{14}$N(n,p)$^{14}$C, respectively.   
The possible production of fluorine via these reactions in massive stars
was first investigated by Woosley \& Weaver (1995), who concluded that the 
above mentioned reactions could not reproduce the abundance of fluorine 
at solar metallicity.
More recently, however, Meynet \& Arnould (2000) investigated 
the role that WR stars may play in the chemical evolution of fluorine by
adopting more modern reaction rates (NACRE compilation) coupled with more
extreme mass loss rates and conclude that production of fluorine
in WR-stars could explain the solar fluorine abundances. 

The crucial point emphasized in the
Meynet \& Arnould (2000) study concerning the chemical evolution
of fluorine, is that any $^{19}$F produced either through the
various proton or $\alpha$-capture reactions must be removed
from the stellar interior to avoid destruction.  In order for 
massive stars
to be significant contributors to net fluorine production, they 
must undergo extensive mass loss; the requirement of large mass-loss rates 
needed to eject the fluorine before its destruction is met by
the WR stars.  Meynet \& Arnould point out
that WR mass loss is very metallicity dependent, and that 
the numbers of WR stars at low metallicities are also very
small.  Thus they predict
that fluorine yields from WR stars will exhibit a strong dependance
on metallicity, with a very small, to negligible yield at low metallicity,
and increasing $^{19}$F yields towards solar-like metallicities.
Our results as shown in Figure 4 with [F/O] plotted versus A(O)
do show a decline in the fluorine to oxygen abundance ratio
as the oxygen abundance decreases, as would be expected 
qualitatively from arguments about the low production of fluorine 
at low metallicities from WR stars.  At this time, it is not possible to
compare quantitatively the observed abundances with predictions
from WR-wind models because there are no published model values of
how the abundance of fluorine varies with oxygen. 

\subsubsection{Massive Stars and the $\nu$-Process}

Perhaps at this time, the leading use for our data on fluorine is as
a test of the proposal that fluorine is primarily
synthesised by the $\nu$-proces in SN II. 
This is possible now because predicted
yields for $^{19}$F (and $^{16}$O, a reference element for SN II
nucleosynthesis) for massive stars of different initial metallicities
were supplied by  Woosley \& Weaver (1995), while Timmes et al.
(1995), Goswami \& Prantzos (2000), and Alib\'{e}s, Labay, \& 
Canal (2001) have incorporated these yields into models of Galactic
chemical evolution, producing predictions of fluorine abundances as
a function of the oxygen abundance.  

The chemical evolution models, with $\nu$-process produced $^{19}$F,
predict that the [F/O] declines fairly steadily as the oxygen
abundance declines; this is illustrated in Figure 4 for two of the
published studies (Timmes et al. 1995 and Alib\'{e}s et al. 2001), 
with both chemical models agreeing quite well.  The rate 
of decrease of [F/O] with A(O) is similar to the results for the
Galactic and LMC stars, with an apparent small offset between model and
observed [F/O].  Recall that there still
remains a small uncertainty in the local Galactic fluorine
abundance, which can be seen in Figure 4 as the offset of the solar
point from the mean of the local K and M giants.  
The slow decline
in [F/O] with A(O) found in the Galactic and LMC stars is fit
well by the $\nu$-process models. 

The consistency between the observed F/O ratios and the model values
is not found for the giants from $\omega$ Cen
for which [F/O] = -0.82 and $\leq$-0.94.
To understand this apparent inconsistency it is
necessary to recognize that $\omega$ Cen is a very different type
of stellar system compared to the LMC or the Milky Way.
Although described as a globular cluster, $\omega$ Cen is far from a
typical Galactic cluster. It is not only the most massive globular
cluster but, in contrast to other clusters, there is a large spread
($\sim$ 1.5 dex) in metallicity amongst its members.
In other
globular clusters (except, perhaps, M 22), the star-to-star spread in
Fe abundance is 0.05 dex or less.  No detailed picture of the formation and
evolution of a globular cluster yet exists but it is thought that
an initial generation of stars polluted 
metal-poor gas from which later generations of stars formed.
There may have been 3 or 4 star-formation episodes in $\omega$ Cen
(Pancino et al. 2000),
although the metal-poor first generation of stars far
outnumber the later generations of more metal-rich stars
(Norris, Freeman, \& Mighell 1996; Suntzeff \& Kraft 1996; 
Norris et al. 1997). 
We suppose that this scenario of a few, well-separated star formation
episodes, which ended many billions of years ago,  
applies to $\omega$ Cen.  This contrasts with the LMC and the Galaxy 
that continue to undergo star formation. 

Consider the following simplest case to describe $\omega$ Cen.  
Ejecta from a first
generation of stars mixes to differing degrees with a reservoir of
very metal-poor gas.
Pockets of gas of differing heavy-element enrichment will result.
Stars comprising a later 
generation that formed from these pockets will have a spread in
heavy-element abundances, as observed in $\omega$ Cen.
Yet, abundance ratios - say, F/O - will be very similar for
all pockets, as long as the primordial gas was severely underabundant in
the elements comprising a ratio.
In the case of an abundance ratio which for SN II ejecta is
dependent on the initial metallicity, the abundance ratio
will stand out as anomalous when judged against the run of the
abundance ratio versus metallicity established from stars that form
from gas that has undergone many episodes of star formation where
the stellar ejecta become well-mixed with the interstellar medium (as
is probably the case for the Galactic and LMC stars).
Abundance ratios insensitive to the initial metallicity
of SN II will appear normal, or nearly so, relative to the Galactic standards.
The most metal-poor stars in $\omega$ Cen have oxygen abundances near
A(O)$\sim$ 7.3 (Smith et al. 2000).  If supernovae of Type II  
with this initial abundance
of oxygen dominated the
nucleosynthesis of fluorine via the $\nu$-process, as predicted by
our simple picture,
then the later generation of $\omega$ Cen
stars would form from gas having [F/O]$\sim$ -0.8 (using $\nu$-process 
model predictions as illustrated in Figure 4). This is confirmed by
the abundance results for the observed $\omega$ Cen stars.

As noted above, our scenario predicts `anomalous' abundance
ratios for all ratios that are dependent on the initial
metallicity of the SN II. Copper is such an example. Significantly,
the [Cu/Fe] from $\omega$ Cen giants (Cunha et al.  2002) is
constant over the [Fe/H] range probed ([Fe/H] = -2 to -0.8) and
falls below its corresponding values in Galactic halo stars. 

\section{Conclusions}

We present the first fluorine abundance measurements outside of
the Milky Way, in the LMC, and the first measurements in a globular 
cluster ($\omega$ Cen).  It is found that the F/O abundance ratios
decline as metallicity (taken as the oxygen abundance) decreases.
This decline is moderate for the LMC (or $\alpha$ Boo) relative to
the near-solar metallicity Galactic disk stars, with the average
value of [F/O] in the LMC being $\sim$ 0.2 dex lower at an oxygen
abundance that is about 0.5 dex lower.  The values of [F/O] are
much lower still in the two $\omega$ Cen giants ([F/O]$\sim$ -0.9)
at a similar oxygen abundance as those in the LMC sample.    
 
The very low values of [F/O] found in the two $\omega$ Cen stars
suggest that AGB stars do not play a dominant role in the
global chemical evolution of fluorine in a stellar population.
Because of the large s-process elemental abundances found in the
more metal-rich $\omega$ Cen stars (as typified by the two $\omega$
Cen targets studied here), a relatively large AGB contribution to
the chemical evolution within $\omega$ Cen is inferred.  If AGB
stars have had such a large impact on the chemical evolution, it
would be expected that this would result in elevated values of [F/O],
instead of the very low values observed.

Both $^{19}$F production sites that involve massive stars, either
via thermonuclear reactions with the synthesized fluorine being
ejected in the high dM/dt stellar winds of a WR star, or from
neutrino spallation off of $^{20}$Ne during core collapse in SN II,
predict lower values of [F/O] at lower oxygen abundances.  With the
results presented here, it is not possible to test conclusively
whether WR winds or the $\nu$-process might dominate the chemical
evolution of $^{19}$F.  This test must await quantitative chemical
evolution models incorporating the yields from WR winds.  In the
meantime, it should be noted that the decline in [F/O] versus A(O)
observed in the Galactic and LMC stars agrees reasonably well with
the chemical evolution models using $\nu$-process yields for
$^{19}$F.

The very low values of [F/O] indicated for the $\omega$ Cen stars
seem, at first glance, at odds with the $\nu$-process models.
This apparent discrepancy is resolved when the very different star
formation history in $\omega$ Cen, when compared to the Milky Way
and the LMC, is considered. 
 
We thank G. Meynet for helpful discussions.
This work is supported in part by the National Science Foundation through
AST99-87374 (VVS) and NASA through NAG5-9213 (VVS). DLL acknowledges
the support of the Robert A. Welch Foundation of Houston, Texas.

\clearpage

\figcaption[fig1.ps]{Observed and synthetic spectra for the star 
LMC 1.27. The synthetic spectra were calculated for A(F) =
3.83, 3.93 and 4.03.  
\label{fig1}}

\figcaption[fig2.ps]{Observed and synthetic spectra for the
$\omega$ Centauri giant ROA324. Three fluorine abundances of
respectively 3.10, 3.4 and 3.98 are presented. An upper
limit fluorine abundance of A(F)= 3.10 was adopted for
this star.
\label{fig2}}

\figcaption[fig3.ps]{The top panel shows the logarithmic abundances
of fluorine plotted versus oxygen.  The solar symbol is shown and the
solid line illustrates a solar F/O abundance ratio.  The bottom panel
is similar to the top, except fluorine is plotted versus iron.  The
A(F) versus A(O) results show slightly less scatter about a solar
F/O line than the A(F) versus A(Fe) values about a solar F/Fe line.
Note the very low F/O ratios in the $\omega$ Cen giants.  
\label{fig3}}

\figcaption[fig5.ps]{Fluorine to oxygen abundance ratios, shown as
[F/O], versus A(O), with the horizontal dashed line depicting a solar
F/O ratio.  The symbols for the various stellar systems are the same
as in Figure 3.  The lines are predictions from chemical evolution models
in which the $\nu$-process contributes to the synthesis of $^{19}$F:
the dashed line is from Timmes et al. (1995) and the solid line is
from Alibes et al. (2001).
\label{fig5}}

\end{document}